\documentclass[12pt]{article}
\usepackage{graphicx}
\usepackage{subfigure}
\usepackage{amsmath}
\usepackage{amssymb}
\usepackage{multicol}
\usepackage{lipsum}



\begin{document}

\begin{center}
\begin{large}
\title\\{ \textbf{An Analysis of Isgur-Wise Function of Heavy-Light Mesons within a Higher Dimensional Potential Model Approach.}}\\\
\end{large}

\author\

\textbf{$Sabyasachi\;Roy^{\emph{a}}\footnotemark\:\:and\:D\:K\:Choudhury^{\emph{a,b,c}}$ } \\\

\footnotetext{Corresponding author. On leave from Karimganj College,Assam, India. e-mail :  \emph{sroy.phys@gmail.com}}
\textbf{a}. Department of Physics, Gauhati University, Guwahati-781014, India.\\
\textbf{b}. Centre for Theoretical Studies, Pandu College, Guwahati-781012, India\\
\textbf{c}. Physics Academy of The North East, Guwahati-781014, India. \\

\begin{abstract}
Nambu-Goto action for bosonic string predicts the quark-antiquark potential to be $V(r) = -\frac{\gamma}{r}+ \sigma r + \mu_0$. The coefficient $\gamma=\frac{\pi(d-2)}{24}$  is the L\"{u}scher coefficient of the L\"{u}scher term  $\frac{\gamma}{r}$, which depends upon the space-time dimension `d'. Very recently, we have developed meson wave functions in higher dimension with this potential from higher dimensional Schrodinger equation by applying quantum mechanical perturbation technique with both $L\ddot{u}scher$ term as parent and as perturbation. In this letter, we analyze Isgur-Wise function for heavy-light mesons using these wave functions in higher dimension and make a comparative study on the status of the perturbation technique in both the cases.

\end{abstract}
\end{center}
\

\begin{multicols}{2}
\section{Introduction:}\rm
In the non-perturbative low energy regime of QCD theory, potential model approach has been successful, even in the non-perturbative approximations, for the study of quark-antiquark bound states. There are several potentials in literature for modeling mesons like Cornell potential [1], Martin potential [2], Logarithmic potential[3], Richardson potential[4] etc. Out of these, the $linear \;plus\; Coulombic$ type Cornell potential is believed to be the more realistic one to account for quark-antiquark interaction in a meson, as it includes both the QCD concepts of asymptotic freedom (Coulombic term) and confinement (linear term).
\begin{equation}
V (r) = -\frac{4\alpha_s}{3r} + br + c
\end{equation}
In the non-relativistic approach, two-body Schr\"{o}dinger equation is employed with such a potential to extract meson wave function and subsequently analysis on static and dynamic properties of mesons are made. The main problem lying within such approach is that Schr\"{o}dinger equation is not exactly solvable with such a potential. In recent past, we have successfully applied quantum mechanical perturbation technique in solving Schr\"{o}dinger equation with such a potential. \\
Very recently,following Nambu-Goto action for bosonic strings[7], we have introduced[5,6] higher dimensional potential model in our studies on heavy-light mesons which is:
\begin{equation}
V ( r ) =-\frac{\gamma}{r}+ \sigma r +\mu_0
\end{equation}
Here, $\gamma=-\frac{\pi(d-2)}{24}$  is the L\"{u}scher coefficient[8,9] of the L\"{u}scher term  $\frac{\gamma}{r}$, which depends upon the space-time dimension 'd'. $\sigma$ is the string tension whose generally accepted value is $0.89 \; GeV/fm$; $\mu_0$ is a regularisation dependent mass term.
We have employed Schrodinger equation in higher dimension and develop meson wave function in both the cases of perturbation - L\"{u}scher term as parent[5] and linear term as parent[6]. Then, we have made subsequent studies on Isgur-Wise function (IWF) of heavy-light mesons in both the cases. \\
In this work, we make some refinement of our analysis of IWF in higher dimension and put forward our comment regarding the status of perturbation for both the choices of `parent and child', within such higher dimensional potential model.
\section{Theory and Calculation:}\rm
\subsection{With L\"{u}scher term as parent:}
With L\"{u}scher term as parent, the wave function is obtained as [5],
\begin{equation}
\Psi^{total}(r)=N_1[1-\frac{\sigma D}{6\gamma}r^{2}]r^{\frac{D-3}{2}}e^{-\mu\gamma r}
\end{equation}
where $D$ is the spatial dimension such that $d=D+1$. $N_1$ is the normalisation constant which is obtained from:
\begin{equation}
\int_0^{\infty} D C_D r^{D-1} \mid \Psi^{total} (r)\mid ^{2}dr =1  \\\
\end{equation}
This gives:
\begin{eqnarray}
N_1=\frac{1}{(D C_D)^{1/2}}.  \nonumber \\
\frac{1}{[\frac{\Gamma(2D-3)}{(2\mu\gamma)^{2D-3}}-2k\frac{\Gamma(2D-1)}{(2\mu\gamma)^{2D-1}}+k^2\frac{\Gamma(2D+1)}{(2\mu\gamma)^{2D+1}}]^{1/2}}
\end{eqnarray}

Here, we consider $k=\frac{\sigma D}{6\gamma}$ and $C_D=\frac{\pi^{D/2}}{\Gamma(\frac{D}{2}+1)}$ such that $d\tau = D C_D r^{D-1} dr$ represents volume element in $D$ dimensional spherical coordinates[10].
It is to be mentioned here that at $D=3$, this higher dimensional wave function gives back our three-dimensional result (see ref [11]) when $\gamma$  is replaced by $\frac{4\alpha_s}{3}$  and $\sigma$ by $b$. Considering volume element in $D$ dimension, the derivatives of IWF are obtained from:
\begin{eqnarray}
\rho^2 = D C_D\mu^2\int_0^\infty r^{D+1}|\Psi^{total}(r)|^2dr \\
C= \frac{1}{6}D C_D\mu^4\int_0^\infty r^{D+3}|\Psi^{total}(r)|^2dr
\end{eqnarray}
With wave function given in (3), we obtain explicit expressions of $\rho^2$ and $C$ :
\begin{flushleft}
\begin{eqnarray}
\rho^{2}=\mu^{2}\frac{[\frac{\Gamma(2D-1)}{(2\mu\gamma)^{2D-1}}-2k\frac{\Gamma(2D+1)}{(2\mu\gamma)^{2D+1}}+k^2\frac{\Gamma(2D+3)}{(2\mu\gamma)^{2D+3}}]}{[\frac{\Gamma(2D-3)}{(2\mu\gamma)^{2D-3}}-2k\frac{\Gamma(2D-1)}{(2\mu\gamma)^{2D-1}}+k^2\frac{\Gamma(2D+1)}{(2\mu\gamma)^{2D+1}}]}\\
C=\frac{\mu^{4}}{6}\frac{[\frac{\Gamma(2D+1)}{(2\mu\gamma)^{2D+1}}-2k\frac{\Gamma(2D+3)}{(2\mu\gamma)^{2D+3}}+k^2\frac{\Gamma(2D+5)}{(2\mu\gamma)^{2D+5}}]}{[\frac{\Gamma(2D-3)}{(2\mu\gamma)^{2D-3}}-2k\frac{\Gamma(2D-1)}{(2\mu\gamma)^{2D-1}}+k^2\frac{\Gamma(2D+1)}{(2\mu\gamma)^{2D+1}}]} \end{eqnarray}
\end{flushleft}
The values of $\rho^2$ and $C$ for different $D$ values are reported in Table-1. For comparison, some recent standard theoretical and experimental results are shown in Table-2.
\subsection{With linear term as parent:}
Considering linear term in potential as parent in perturbation method, the wave function comes out as:
\begin{eqnarray}
\Psi^{total}(r)= N_1r^{\frac{(1-D)}{2}}[1+A_1(r,D)r+A_2(r,D)r^{2}+ \nonumber \\
 A_3(r,D)r^{3}+.........](\varrho_1 r)^{m}Ai[\varrho_1 r-\varrho_0]
\end{eqnarray}
Terms $A_1, A_2$ etc, Airy function $Ai[\varrho]$ and other terms involved in wave function are described in ref [6]. Normalisation constant $N_1$ and slope ($\rho^2$) and curvature ($C$) can be obtained using equations (4), (6) and (7) respectively. To avoid divergences due to infinite upper limit of integration in these expressions, from convergence condition of perturbation series, we have earlier introduced reasonable cut-off value $r_0$, which is obviously dimension dependent (Table-1  of ref[6]). The variation of $\rho^2$ and $C$ with dimension $D$ is reported in Table-3 of ref [6], which we also report here as Table-3, for convenience of our analysis.
\section{Results and discussion:}\rm
In the case of L\"{u}scher term as parent, although our results ( equations (3), (8), (9)) for $D=3$ give back the corresponding expressions obtained with Cornell potential (see[11])when $\gamma$ is replaced by $\frac{4\alpha_s}{3}$ and $\sigma$ by $b$, still our result for $\rho^2$ and $C$ ( Table-1) are higher than the standard values. \\
However, at higher $D$, when L\"{u}scher term becomes more and more dominant, the values of $\rho^2$ and $C$ go on decreasing, the asymptotic limits being $\rho^2_{asym}= 15.32 $ and $C = 36.64 $ for D meson.  Here, we mention that, we have worked with fixed value of confinement parameter($\sigma =0.89 \;GeV/fm $) , unlike earlier analysis with Cornell potential (see [11]) where confinement parameter $b$ is varied suitably to meet the expectations. Also, the Coulomb-type L\"{u}scher term depends only upon dimensional parameter $D$, where as in Cornell potential the coefficient of Coulombic term contains strong coupling constant $\alpha_s$, bringing in flexibility in the calculations. \\
On the other hand, with linear term as parent in perturbation technique, our results (Table-3) are reasonably close to the standard values ( Table-2). More importantly, in this case, $\rho^2$ and $C$ values increase with increase in $D$,although remaining within the range of our expectations. We have restricted our calculation up to the term $A_5(r,D)$ in the first infinite series of the wave function; there remains scope for further refinement of results.\\
From our analysis of derivatives of IWF of heavy-light mesons in both the choices of `parent-child', we can infer that the case of linear confinement term as parent is giving more reliable results compared to the case of L\"{u}scher term as parent. This is also supported by recent lattice calculation [14]. Also, theoretically it is reasonable that the higher dimensional potential originates from a more general string inspired exact potential enunciated by J. F. Arvis[15].
\begin{equation}
V(r)=\sigma r \sqrt{1-\frac{\pi(d-2)}{12\sigma r^2}}
\end{equation}
On expansion, it gives linear term as leading one with L\"{u}scher term as first order correction to it. This supports the case of treating linear term as parent with L\"{u}scher term as first order perturbation to it.
\paragraph{Acknowledgement:}
\begin{flushleft}
$SR$ acknowledges the support of University Grants Commission, Govt. of India in terms of fellowship under $FDP$ scheme to pursue his research work at Gauhati University.
\end{flushleft}

\end{multicols}

\begin{table*}[htb]
\begin{center}
\caption{$\rho^{2}$ and $C $ for different $D$ values (with L\"{u}scher parent).}\label{centre}
\begin{tabular}{|c|cc|cc|}
  \hline
   &  D meson & & B meson & \\
     \hline \hline
       $D$ &  $\rho^2$ & C   & $\rho^2$ & C   \\
   \hline
   3  & 186.1 & 11200.7 & 186.4 &  11208.8  \\    \hline
   4  & 76.4  & 2032.1  & 76.9  &  2051.6   \\    \hline
   5  & 50.51 & 854.9   & 50.82 &  563.7    \\    \hline
   9  & 29.39 & 204.2   & 29.53 &  208.8    \\    \hline
   15 & 26.93 & 165.4   & 27.21 &  169.2    \\    \hline
   20 & 22.41 & 96.6    & 22.68 &  99.2     \\    \hline
   25 & 18.83 & 64.23   & 19.02 &  64.6     \\    \hline
   $\infty$ & 15.32 & 36.64 & 15.55 & 36.98 \\
   (asym) &&&& \\ \hline
\end{tabular}
\end{center}
\end{table*}
\begin{table}[!htbp]
\begin{center}
\caption{$\rho^{2}$ and $C $ values in different models/collaborations.}\label{cross}
\begin{tabular}{|c|r|r|}
  \hline
Model / collaboration &	Slope & Curvature \\
\hline \hline
 Ref [16] & 1.8 & 0.95 \\
 Skryme Model [17] & 1.3 & 0.85 \\
 Neubert [18] & 0.82   & -- \\
 UK QCD Collab. [19]  & 0.83 & -- \\
 CLEO [20] & 1.67 & -- \\
 BELLE  [21] & 1.35 & -- \\
HFAG [22] & 1.17 $\pm 0.05$ & -- \\
Huang [23] & 1.35 $\pm 0.12$ & -- \\
  \hline
\end{tabular}
\end{center}
\end{table}
\begin{table}[!htbp]
\begin{center}
\caption{$\rho^{2}$ and $C $ for different $D$ values (with linear parent).}\label{centre}
\begin{tabular}{|c|cc|cc|}
  \hline
   &  D meson & & B meson & \\
     \hline \hline
       $D$ &  $\rho^2$ & C   & $\rho^2$ & C   \\
   \hline
   3  & 0.2158 & 0.0174 & 0.2608 & 0.0254    \\    \hline
   4  & 0.3874 & 0.0457 & 0.4039 & 0.0483    \\    \hline
   5  & 0.4366 & 0.0613 & 0.4813 & 0.0679    \\    \hline
   9  & 0.8355 & 0.1746 & 0.9497 & 0.1916    \\    \hline
   15 & 0.9622 & 0.2710 & 1.0281 & 0.3274    \\    \hline
   20 & 1.2178 & 0.3997 & 1.3352 & 0.4887    \\    \hline
   25 & 1.3732 & 0.5041 & 1.4637 & 0.6623    \\    \hline
\end{tabular}
\end{center}
\end{table}


\begin{thebibliography}{99}


\bibitem{1}  E. Eicheten \emph{et al.}, Phy.Rev.\textbf{D17}, 3090(1978); Phy.Rev. \textbf{D21},203(1980).
\bibitem{2}	 A. Martin, Phy. Lett. \textbf{B93},338(1980).
\bibitem{3}  C. Quigg and J. L. Rosuer, Phy.Lett.\textbf{B 71},153(1977).
\bibitem{4}  J. L.Richardson, Phy. Lett. \textbf{B82}, 272(1979).
\bibitem{5}  S Roy, B J Hazarika and D K Choudhury, Phys. Scr. \textbf{86},045101(2012).
\bibitem{6}  S. Roy and D. K. Choudhury, arXiv: 1301.0982 [hep-ph].
\bibitem{7}  Y. Nambu, Phy. Lett. \textbf{B80}, 372(1979).
\bibitem{8}  M. L\"{u}scher,K. Symanzik and P.Weisz, Nucl.Phys. \textbf{B173}, 365(1980).
\bibitem{9}  M. L\"{u}scher, Nucl. Phy. \textbf{B 180}, 317(1981).
\bibitem{10} Noura Eduarda, David G.Henderson in ``Experiencing Geometry: On plane and sphere", Prentice Hall(1996).
\bibitem{11} N. S.Bordoloi and D. K. Choudhury, IJMP A,Vol 15 No.23, 3667(2000).
\bibitem{12} N. Isgur and M. B. Wise, Phy. Rev. Lett. \textbf{66}, 1130(1991).
\bibitem{13} S. Roy, N S Bordoloi and D. K. Choudhury, Can.J.Phy \textbf{91}, 34(2013).
\bibitem{14} G.S. Bali, Phys. Lett. \textbf{B460}, 170(1999)[hep-ph/9905387].
\bibitem{15} J. F. Arvis, Phy.Lett.\textbf{B127}, 106(1983).
\bibitem{16}  S. Roy and D. K. Choudhury, Mod. Phy. Lett. \textbf{A27}, 1250110 (2012).
\bibitem{17}	E. Jenkins, A. Manahar, M. B. Wise; Nucl. Phys B , 396; 38(1996).
\bibitem{18}	M. Neubert, Phys Lett B 264; 455(1991).
\bibitem{19}	UKQCD Collab. , K C Bowler et al; Nucl Phys B, 637, 293(2002).
\bibitem{20}	CLEO Collab., J Bartel et al, Phys Rev Lett 82, 3746(1999).
\bibitem{21}	BELLE collab, K Abe et al, Phys Lett B, 526, 258(2002).
\bibitem{22}	Heavy Flavor Averaging Group (HFAG), hep-ex/08081297(2009).
\bibitem{23}	Ming-Qiu Huang et al, Phys.Letts. B 629,27-32(2005).
\end{thebibliography}
\end{document}